\documentclass[prl,superscriptaddress,twocolumn]{revtex4-1}
\usepackage{amsmath,mathrsfs}
\usepackage{graphicx}
\usepackage{rotating}
\usepackage{amssymb}

\usepackage[usenames, 
            dvipsnames
            ]{color}

\usepackage[colorlinks=true,
            linkcolor=Blue,
            filecolor=magenta,
            citecolor=Blue,
            urlcolor=Blue,
            pdftitle={Dipolar Spins},
            pdfauthor={A. Keles},
            bookmarks=true
            ]{hyperref}

\usepackage[sort&compress]{natbib}

\usepackage{float}

\usepackage[numbered, 
            open
            ]{bookmark}

\usepackage[all]{hypcap}

\graphicspath{ {./figs/} }

\usepackage{listings}

\usepackage{tikz}

\tikzset{middlearrow/.style={
    decoration={markings,
      mark= at position 0.55 with {\arrow[scale=1,blue]{#1}} ,
    },
    postaction={decorate}
  }
}
\usetikzlibrary{decorations.pathreplacing,
  decorations.markings,decorations.pathmorphing}
\usetikzlibrary{calc}

\usepackage{multirow}

\usepackage{pifont}

\newcommand\dgr{\ensuremath^\dagger}

\newcommand\w{\ensuremath\omega}
\newcommand\x{\ensuremath\mathbf{x}}
\newcommand\rb{\ensuremath\mathbf{r}}

\newcommand\pb{\ensuremath\mathbf{p}}
\newcommand\qb{\ensuremath\mathbf{q}}

\usepackage{bbold}

\usepackage[normalem]{ulem}

\newcommand\kelesout{\bgroup\markoverwith{\textcolor{red}{\rule[0.5ex]{2pt}{1.1pt}}}\ULon}

\newcommand{\oneloop}[1]
{
\begin{tikzpicture}[scale=.4,baseline=(current bounding box.center)]

    \def \x {0}
    \def \y {.1}
    \def \w {.3}
    \def \l {.5}  
    \def \dver {3}
    \draw[fill=gray] (\x,\y) rectangle (\x+\w,\y+\w);
    \draw[fill=gray] (\x+\w+\dver,\y) rectangle (\x+\dver+2*\w,\y+\w);
    \foreach \m/\n [count=\i] in {#1}
    {
      \ifthenelse{\equal{\i}{1}}
      {
        \ifx\m\n
          \draw[middlearrow={stealth}] 
        \else
          \draw[middlearrow={stealth reversed}] 
        \fi
        (\x-\l,\y+\w+\l) -- (\x,\y+\w);
        \node [left, black] at (\x-\l,\y+\w+\l) 
        {\tiny\ensuremath{\m}};
      }{}  
      \ifthenelse{\equal{\i}{2}}
      {
        \ifx\m\n
        \draw[middlearrow={stealth}] 
        \else
        \draw[middlearrow={stealth reversed}] 
        \fi
        (\x-\l,\y-\l) -- (\x,\y);
        \node [left, black] at (\x-\l,\y-\l) 
        {\tiny\ensuremath{\m}};
      }{} 
      \ifthenelse{\equal{\i}{3}}
      {
        \ifx\m\n
        \draw[middlearrow={stealth}] 
        \else
        \draw[middlearrow={stealth reversed}] 
        \fi
        (\x+2*\w+\dver,\y+\w) -- (\x+2*\w+\dver+\l,\y+\w+\l);
        \node [right, black] at (\x+2*\w+\dver+\l,\y+\w+\l) 
        {\tiny\ensuremath{\m}};
      }{} 
      \ifthenelse{\equal{\i}{4}}
      {
        \ifx\m\n
        \draw[middlearrow={stealth}] 
        \else
        \draw[middlearrow={stealth reversed}] 
        \fi
        (\x+2*\w+\dver,\y) -- (\x+2*\w+\dver+\l,\y-\l);
        \node [right, black] at (\x+2*\w+\dver+\l,\y-\l) 
        {\tiny\ensuremath{\m}};
      }{} 
      \ifthenelse{\equal{\i}{5}}
      {
        \ifx\m\n
        \draw[middlearrow={stealth reversed}] 
        \else
        \draw[middlearrow={stealth }] 
        \fi
        (\x+\w+\dver,\y+\w) 
        to [out=145,in=35]
        (\x+\w,\y+\w);
        \node [above, black] at (\x+\w+.5*\dver,\y+\w+.5)
        {\tiny\ensuremath{\m}};
      }{} 
      \ifthenelse{\equal{\i}{6}}
      {
        \ifx\m\n
        \draw[middlearrow={stealth reversed}] 
        \else
        \draw[middlearrow={stealth }] 
        \fi
        (\x+\w+\dver,\y) 
        to [out=-145,in=-35] 
        (\x+\w,\y);
        \node [below, black] at (\x+\w+.5*\dver,\y-0.5) 
        {\tiny\ensuremath{\m}};
      }{} 
    } 
  \end{tikzpicture}
}

\begin{document}
\title{Absence of long-range order in a triangular spin system with dipolar
interactions}
\author{Ahmet Kele\c{s}}
\affiliation{Department of Physics and Astronomy,
  University of Pittsburgh, Pittsburgh, Pennsylvania 15260, USA}
\affiliation{Department of Physics and Astronomy,
  George Mason University, Fairfax, Virginia 22030, USA}
\author{Erhai Zhao}
\affiliation{Department of Physics and Astronomy,
  George Mason University, Fairfax, Virginia 22030, USA}

\begin{abstract}
Antiferromagnetic Heisenberg model on the triangular lattice is perhaps the
best known example of frustrated magnets, but it orders at low temperatures.
Recent density matrix renormalization group
(DMRG) calculations find that next nearest neighbor interaction $J_2$ 
enhances the frustration and leads to a spin liquid for $J_2/J_1\in (0.08,0.15)$. 
In addition, DMRG study of a dipolar Heisenberg model with longer range interactions gives 
evidence for a spin liquid at small dipole titling angle $\theta\in
[0,10^\circ)$. In both cases, the putative spin liquid region
appears to be small.
%
Here, we show that for the triangular lattice dipolar Heisenberg model, a robust quantum 
paramagnetic phase exists in a surprisingly wide region, $\theta\in [0,54^\circ)$, for dipoles tilted along the lattice diagonal direction.
We obtain the phase diagram of the model by functional renormalization
group (RG) which treats all magnetic instabilities on equal footing. The quantum paramagnetic phase
is characterized by a smooth continuous flow of vertex functions and spin susceptibility down to the lowest RG scale, in contrast
to the apparent breakdown of RG flow in phases with stripe or spiral order. Our finding points to a promising direction 
to search for quantum spin liquids in ultracold
dipolar molecules.
\end{abstract} 

\maketitle

Quantum spin liquids evade conventional long-range order or symmetry breaking
down to zero temperature \cite{ balents2010spin, lee2014quantum,
  savary2016quantum, RevModPhys.89.025003}.  These highly entangled states
have unique properties including possible topological order or fractional
excitations. Theoretically, the existence of certain spin liquid states is
firmly established from exactly solvable models
\cite{kitaev2003fault,kitaev2006anyons}. While powerful numerical methods such
as Density Matrix Renormalization Group (DMRG) and tensor networks (TN) have
yielded clear evidence for spin liquids in geometrically frustrated spin
models including the kagome lattice spin $1/2$ Heisenberg model
\cite{yan2011spin,PhysRevLett.109.067201} and  the triangular lattice
$J_1$-$J_2$ Heisenberg model \cite{PhysRevB.92.041105,dsheng}, the very nature
of these spin liquids remains controversial.  Experimentally, two class of
materials, herbertsmithite \cite{shores2005structurally} with kagome lattice
structure and triangular lattice organic compounds
\cite{PhysRevLett.91.107001,PhysRevLett.95.177001,yamashita2010highly}, have
emerged as strong candidates for quantum spin liquids. In the continuing
search for spin liquids, it is useful to examine other model spin systems that
are experimentally accessible. 

A new class of quantum spin models, dubbed dipolar Heisenberg model, with
long-range exchange interactions were recently predicted to harbor spin
liquids. This model can be realized  using polar molecules confined in deep
optical lattices
\cite{yan2013observation,PhysRevLett.113.195302,PhysRevA.84.033619,yao2015quantum}.
Similar spin models with tunable range and anisotropy have also been
experimentally demonstrated with cold atoms with large magnetic moments
\cite{PhysRevLett.111.185305}, Rydberg-dressed atoms
\cite{schauss2012observation,labuhn2016tunable}, and trapped ions
\cite{britton2012engineered,islam2013emergence}. These experiments thus
motivate the exploration of the phase diagrams of dipolar Heisenberg model.
Compared to the $J_1$-$J_2$ model, further range exchanges compete and
sometimes enhance frustration.  For example, TN calculation shows a narrow
region of paramagnetic phase on the square lattice
\cite{PhysRevLett.119.050401} which is also supported by RG analysis \cite{keles2018renormalization}. 
In Ref. \onlinecite{yao2015quantum}, DMRG
predicts a spin liquid phase on the triangular lattice for $\theta$ between 0
to 10 degrees, where the dipole tilting angle $\theta$ controls the spatial
anisotropy of the exchange.  The spin liquid regions however seem small for
both lattices. In addition, both DMRG and TN are limited to small lattice
sizes: the range of interaction has to be truncated, and a small cluster is
insufficient to accommodate the spiral order which has incommensurate wave
vector and occupies much of the classical phase diagram. An independent,
alternative approach is needed.

\begin{figure}[h]
  \includegraphics[width=0.43\textwidth]{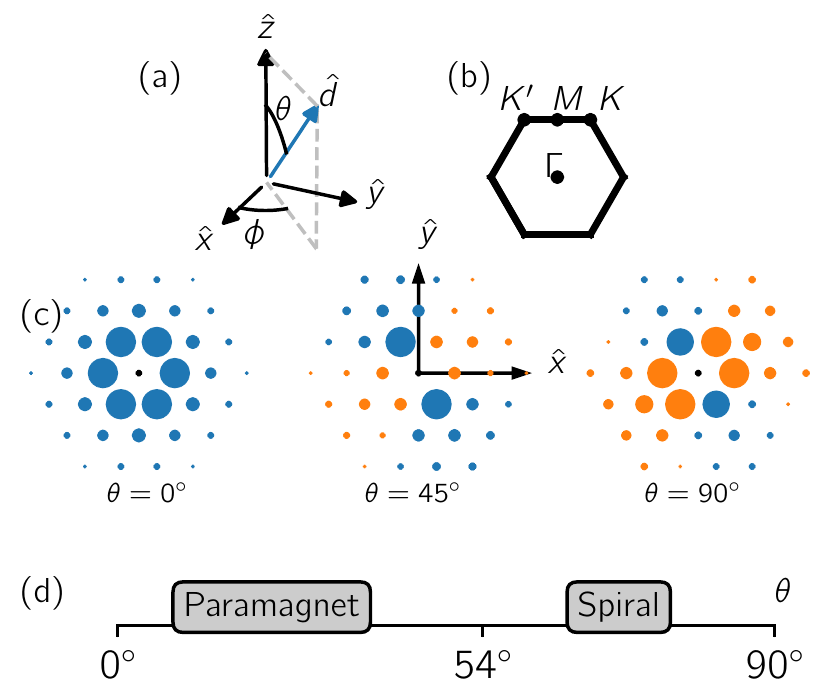}
  \caption{(Color online) Dipolar Heisenberg model on the triangular lattice
    in the $xy$ plane. (a) The exchange $J_{ij}$ depends on the dipole
    orientation $\hat{d}$, with polar angle $\theta$ and azimuthal angle
    $\phi$. (b) High symmetry points within the Brillouin zone. (c) The
    competing exchange couplings $J_{ij}$ (site $i$ is at the origin) for
    $\phi=30^\circ$ and $\theta=0^\circ$, $45^\circ$ and $90^\circ$
    respectively. The size of the circle indicates the magnitude of $J_{ij}$,
    with positive (negative) $J_{ij}$ shown in blue (orange). (d) The zero
    temperature phase diagram at $\phi=30^\circ$ includes a wide quantum
    paramagnetic phase, $\theta\in [0,54^\circ)$, and a spiral phase.}
  \label{fig:geometry}
\end{figure}

\begin{figure}
  \vspace{0.1in}  
  \centering
  \includegraphics[scale=0.9]{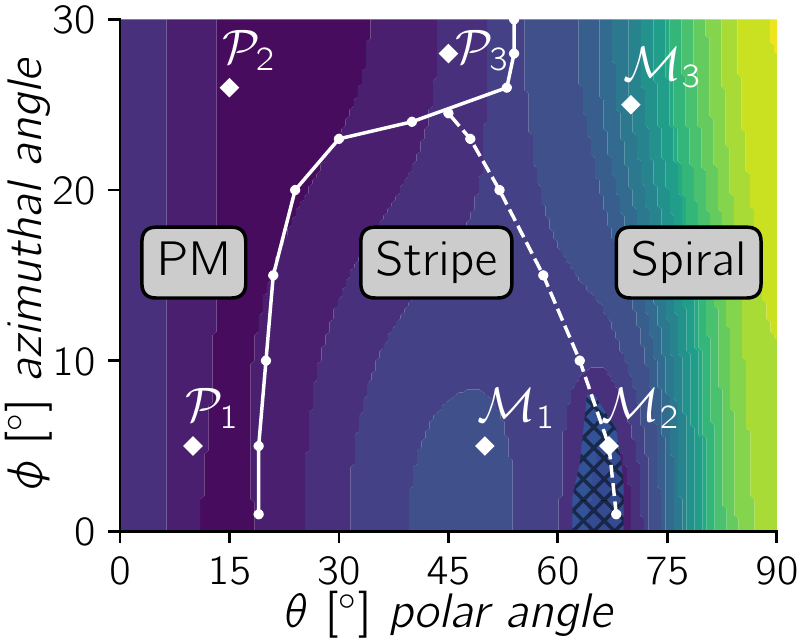}
  \includegraphics[scale=0.9]{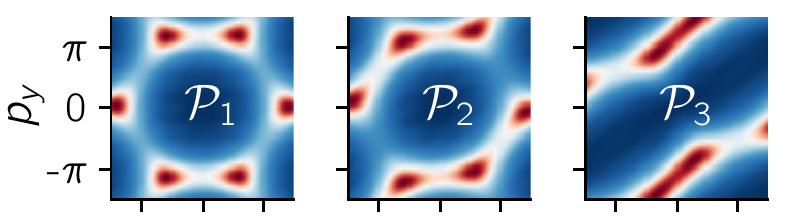}
  \includegraphics[scale=0.9]{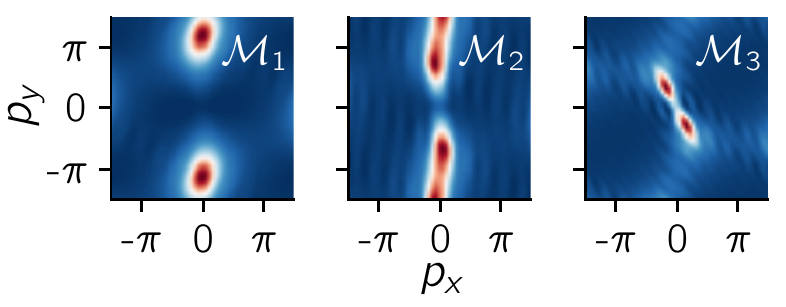}
  \includegraphics[scale=0.9]{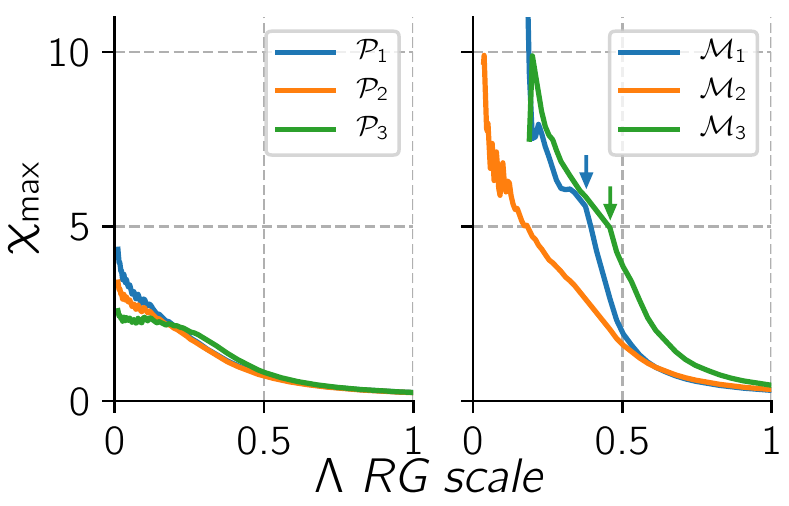}
  \caption{(Color online) Zero temperature phase diagram of the dipolar Heisenberg model on
    triangular lattice as a function of the dipole tilting angle ($\theta$,
    $\phi$) showing the quantum paramagnetic (PM), stripe and spiral
    phases. Solid and dashed white lines are the phase boundaries (see main
    text).  Color contours shows estimated critical temperature $\Lambda_s$
    from static FRG. Spin susceptibility profiles $\chi(\mathbf{p})$ at scale
    $\Lambda<0.5$ are shown in the middle panels for representative points 
    marked with $\mathcal{P}_1$, $\mathcal{M}_1$ etc.. The
    flows of peak susceptibility $\chi_\mathrm{max}$ are shown in the bottom
    panels. For $\mathcal{M}_1$ and $\mathcal{M}_3$, the flow becomes unstable
    below some RG scale indicated by the arrows. }
  \label{fig:triangular}
\end{figure}

In this paper, we provide compelling evidence that the spin liquid region of
the dipolar Heisenberg model can be expanded by five fold, to
$\theta\in[0,54^\circ)$, by tilting the dipoles towards the diagonal of the
triangular lattice. Our idea exploits the tunable anisotropy available in
experiments to suppress the stripe phase to arrive at a 
simple phase diagram which contains the quantum paramagnetic phase and the
spiral phase, see Fig. \ref{fig:geometry}(d). We argue that quantum
paramagnetic phase is a spin liquid by comparing to DMRG. We further
obtain the full phase 
diagram for arbitrary dipole tilting (Fig. \ref{fig:triangular}) using numerical functional
renormalization group which is capable of handling long-range interactions and
spiral order using large cutoffs for the interaction range. 

{\it Dipolar Heisenberg model and its classical phases.} Consider dipolar
molecules localized in a deep optical lattice. Two rotational states of the
molecule can be isolated to play the role of pseudospin 1/2. The dipole-dipole
interaction induces long-range exchange interactions of the Heisenberg form
(see Ref. \cite{PhysRevA.84.033619,yao2015quantum,PhysRevLett.119.050401}
for details)
\begin{equation}
    H = \sum_{i\neq j} J_{ij} \mathbf{S}_i\cdot \mathbf{S}_j ,
    \label{eq:hamiltonian}
\end{equation}
where the sum is over {all} pairs of sites in a triangular
lattice and
$\mathbf{S}_i=(S_i^x,S_i^y,S_i^z)$ are spin half
operators at site $i$.  We assume one molecule per site and all the dipole moments 
oriented along a common direction $\hat d$ set by an external electric field. In terms of the 
polar angle $\theta$ and the azimuthal angle $\phi$ as shown in
Fig.~\ref{fig:geometry}(a), $\hat d = (\sin\theta\cos\phi, \sin\theta\sin\phi, \cos\theta)$. The
exchange interaction then takes the dipolar form 
\begin{equation}
    J_{ij} = J_0[1-3(\hat r_{ij}\cdot \hat d)^2 ]/r_{ij}^3 
    \label{eq:dipolar_J}
\end{equation}
where $\mathbf{r}_{ij}=\mathbf{r}_i-\mathbf{r}_j$ for spins at sites
$\mathbf{r}_i$ and $\mathbf{r}_j$. Here the lattice constant is taken to be
unity and the energy unit is given by $J_0$, the leading dipolar exchange.

The dipolar Heisenberg model, Eq. \eqref{eq:hamiltonian}-\eqref{eq:dipolar_J},
is severely frustrated. In addition to the lattice geometric frustration, 
various (the nearest, second and further neighbor) exchanges, 
with their relative magnitude and sign controlled by dipole tilting
[Fig.~\ref{fig:geometry}(c)], prefer different, competing 
long-range orders.  
To appreciate the possible orders, we first solve this model for classical
spins \cite{supp}. Consider for example the case of $\phi=0$, i.e. $\hat{d}$ titling
along the $x$-axis, and varying $\theta$. 
From $\theta=0^\circ$ to $\sim 20^\circ$, it has the familiar 120$^\circ$ order. Stripe order takes over
for $\theta\in (20^\circ,60^\circ)$, with the spins aligned along $x$ but alternating ($\mathbf{S}\rightarrow -\mathbf{S}$)
along $y$. For all other $\theta$ values, the classical ground
state is a spiral with incommensurate wave vector $\mathbf{q}(\theta,\phi)$. 
As $\phi$ is increased, the stripe phase shrinks and eventually vanishes.
It is largely dictated by symmetry:
stripes along the lattice direction would break the reflection symmetry of the Hamiltonian
with respect to the $\hat{z}-\hat{d}$ plane and cost energy. This trend will continue to hold
in the quantum phase diagram. The energy minima of the spiral and 120$^\circ$ phase are 
very shallow, a symptom of frustration \cite{supp}. As we will show below, they are easily melted by quantum fluctuations, 
leading to a drastically reconstructed phase diagram, Fig.~\ref{fig:geometry}(d).

{\it Pseudo-fermion FRG.} The key to find the phase diagram of the quantum dipolar Heisenberg model
is an accurate, unbiased many-body technique that can treat the spiral order, long-range interactions,
and large lattices.
Functional renormalization group (FRG) is well suited for this purpose. It starts with the bare interaction, and 
systematically integrates out the high energy, short wave length fluctuations to track the flow of 
the effective action functional. Under the flow toward lower energy and longer wave length, the leading many-body instability emerges as the dominant divergence.
We follow the pseudo-fermion FRG (pf-FRG)
put forward by Reuther and W\"olfle, which has been extensively benchmarked against other methods and  
applied to frustrated spin models
\cite{Reuther2010,Reuther2011,Reuther2011b,Reuther2014,PhysRevB.89.020408,Buessen2016,PhysRevB.92.220404,Iqbal2016}.
The spin model Eq.~\eqref{eq:hamiltonian} is first rewritten
in a fermionic representation via 
$\mathbf{S}_i=\frac{1}{2}\mathbf{\sigma}_{\alpha\beta} \psi\dgr_{\alpha i}\psi_{\beta
  i}$ where $\psi$'s are fermionic field operators. The resulting interacting fermion
  problem is then solved using the well-established fermionic FRG developed for 
strongly correlated electrons \cite{Metzner2012,P.KopietzL.BartoschSchutz2010,bhongale2012bond,kelecs2016competing}.
Specifically, vertex expansion up to one loop order yields the flow equations for 
the fermion self energy $\Sigma$ and the effective interaction vertex $\Gamma$ as functions of the
sliding RG scale $\Lambda$,
\begin{align}
    \partial _{\Lambda}
    \Sigma( \omega_1 ) 
    &=-\sum_2 \Gamma_{1,2;1,2} S(\omega_2)  ,
    \label{eq:self_energy_flow}
    \\
    \partial _{\Lambda}
    \Gamma_{ 1',2';1,2 } 
    &= \sum_{3,4} \Pi(\omega_3,\omega_4 ) 
    \biggr[
     \frac{1}{2} \Gamma_{ 1',2';3 ,4 }\Gamma_{ 3 ,4 ;1 ,2 } 
     \nonumber\\
    &-\Gamma_{ 1',4 ;1 ,3 }\Gamma_{ 3 ,2';4 ,2 } 
    +\Gamma_{ 2',4 ;1 ,3 }\Gamma_{ 3 ,1';4 ,2 }
    \biggr].
    \label{eq:vertex_flow}
\end{align}
Hereafter the $\Lambda$ dependence of $\Sigma$, $\Gamma$, $G$, $S$ etc.
is omitted for brevity, and we use the shorthand notation $\Gamma_{1',2';1,2}\equiv\Gamma(
i_1,\alpha_1,\omega_1,i_2,\alpha_2,\omega_2;
i_1',\alpha_1',\omega_1',i_2',\alpha_2',\omega_2' )$ with site index $i$,
spin $\alpha$ and frequency $\omega$. The sum
denotes integration over $\omega$ as well as summation over lattice
sites and spin. The scale-dependent propagators are defined by
\begin{equation}
G(\omega)=\frac{\Theta(|\omega|-\Lambda)}{i\omega+\Sigma(\omega)},\quad
S(\omega)=\frac{\delta(|\omega|-\Lambda)}{i\omega+\Sigma(\omega)}.\quad
\end{equation}
Note that the bare fermion propagator only has frequency dependence, 
$G^{(0)} (\omega) = 1/i\omega$
\footnote{
The chemical potential
is set to zero to enforce the fermion number constraint, see Ref. \cite{Reuther2010}.}. 
Eq.~\eqref{eq:vertex_flow} includes the
particle-particle, the particle-hole as well as the exchange channel as shown by the
following diagrams:
\begin{equation}
    \oneloop{1,2,1',2',3,4}+
\oneloop{1,1'/,2/,2',3,4/}+
\oneloop{1,2'/,2/,1',3,4/}
\nonumber
\end{equation}
We adopt an improved truncation scheme beyond one loop 
\cite{PhysRevB.70.115109} where the bubble $ \Pi$ is given by the {full} derivative
\begin{equation}
 \Pi(\omega_3,\omega_4 ) =-\frac{d}{d\Lambda}
 \left[
 G(\omega_3)G(\omega_4)
 \right].
\end{equation}
The first order nonlinear integro-differential equations in
Eq.~\eqref{eq:self_energy_flow} and \eqref{eq:vertex_flow} are supplemented
by the following initial conditions at the ultraviolet scale
$\Lambda_\mathrm{UV}\rightarrow \infty$,
\begin{align}
  \Sigma(\omega)|_{\Lambda_\mathrm{UV}}   &= 0,
  \label{eq:initial_condition}\\
  \Gamma_{1,2;1',2'}|_{\Lambda_\mathrm{UV}} &= \frac{1}{4}
  \sigma^\mu_{\alpha_1\alpha_1'} \sigma^\mu_{\alpha_2\alpha_2'}
  J_{i_1,i_2}\delta_{i_1i_1'}\delta_{i_2i_2'} - (1'\leftrightarrow 2').  
    \nonumber
\end{align}

We numerically solve the coupled flow equations Eq.
\eqref{eq:self_energy_flow}-\eqref{eq:vertex_flow} together with the initial
condition Eq. \eqref{eq:initial_condition} and the dipolar exchange Eq.
\eqref{eq:dipolar_J} using the fourth order Runge-Kutta, for  a logarithmic frequency grid of
$N_\omega$ frequencies, by taking $N_\Lambda$ RG steps from bare scale
$\Lambda_\mathrm{UV}$ down to zero. We keep all couplings $\Gamma$ within an $N_L\times N_L$
parallelogram on the triangular lattice \cite{supp}. 
The computational cost scales with $N_\Lambda\cdot N_L^2\cdot N_\omega^4$. We perform simulations up
to $N_L=13$, $N_\omega=64$, and $N_\Lambda=4N_\omega$, i.e. four RG steps between
two neighboring frequency points. Following the efficient spin and frequency
parametrization scheme of Ref. \onlinecite{Reuther2010}, and exploiting the
reflection symmetry of $H$, we still end up with over 22 million coupling
constants ($\Gamma$s).  To make the calculation tractable, the FRG code is
designed to run parallel on thousands of graphic processing units. We have
benchmarked it and found good agreement with known FRG results on the square \cite{Reuther2010}
and triangular \cite{Iqbal2016} lattice $J_1$-$J_2$ model.

From $\Gamma$ and $\Sigma$, we compute the static spin-spin correlation
functions in real space and then Fourier transform to obtain the spin
susceptibility $\chi(\mathbf{p})$ \cite{Reuther2010}. Let $\chi_\mathrm{max}$ be the maximum
value of $\chi$ reached at wave vector $\mathbf{p}=\mathbf{p}_\mathrm{max}$
within the Brillouin zone shown in Fig.~\ref{fig:geometry}(b).  Together $\chi(\mathbf{p})$ and $\mathbf{p}_\mathrm{max}$ 
offer clues about the onset or lack
of long-range order under the FRG flow. Typically $\chi_\mathrm{max}$
displays Curie-Weiss behavior for $\Lambda \gg 1$ until the build-up of
quantum correlations starts to kick in around $\Lambda \sim 1$. 
An instability towards long range order is signaled by the divergence of
$\chi_\mathrm{max}$ at some critical scale $\Lambda_c<1$. The finite cluster
size and the truncation and discretization
regularize the divergence, and replace it with unstable, irregular and oscillatory flow below $\Lambda_c$.  Despite this, the breakdown
of the smooth flow is unmistakable, and the type of incipient order can be
inferred from the location of $\mathbf{p}_\mathrm{max}$. 
It may also happen that the flow of $\chi$ remains stable and smooth down to
the lowest RG scale $\Lambda\rightarrow 0$. Then the system settles into a
paramagnetic phase.

To orient the full FRG solution and compare with the classical results, we
first carry out static FRG, i.e. solving the flow equations by ignoring all
$\omega$ dependences \cite{supp, PhysRevB.96.045144}. This approximation was shown
to be consistent with random phase approximation and Luttinger-Tisza method
\cite{PhysRevB.96.045144}. From the flow of $\Gamma$, we extract a ``critical
scale" $\Lambda_s$, at which the maximum value of $\Gamma$ reaches a large
cutoff value (diverges). Thus $\Lambda_s$ serves as a rough estimation of the
critical temperature for the long-range order.  Fig.~\ref{fig:triangular}
shows the resulting $\Lambda_s$ in false color with contour lines.  Here we
find good agreement with the classical analysis. The $120^\circ$ order, where
$\chi$ shows maxima at the corners of the Brillouin zone, lies at small
$\theta$. For increasing $\theta$, peaks at $K$ and $K'$ come together and
merge at the $M$ point, indicating the stripe phase. For even larger
$\theta$, the peak at $M$ moves towards the $\Gamma$ point, signaling the
spiral order.  Fig.~\ref{fig:triangular} also shows that the spiral phase has
the largest $\Lambda_s$ (in green) whereas $\Lambda_s$ is significantly
suppressed (in dark blue) in the region around $\theta\sim 15^\circ$ and near
the phase boundaries. These dark areas are where spin liquid is suspected to
reside.

{\it Phase diagram from pf-FRG.} Solving the flow equations with full 
frequency dependence along multiple cuts on the $(\theta,\phi)$ plane
and examining the flow of $\chi_\mathrm{max}$ and the profile of 
$\chi(\mathbf{p})$ in momentum space, we arrive at 
the zero temperature phase diagram of the dipolar Heisenberg model 
on triangular lattice shown in Fig.~\ref{fig:triangular}. The solid
line and the dashed line mark the phase boundary between 
three major phases: the stripe, the spiral, and the quantum paramagnetic
(PM) phase. The most striking result from pf-FRG is {\it the abundance of 
the PM phase}. A large portion of the classical spiral, including the $120^\circ$ order, 
completely melts due to strong quantum fluctuations and gives way to quantum paramagnet.
Compared to the $J_1$-$J_2$ Heisenberg model which shows a narrow region of spin
liquid between the $120^\circ$ and stripe order \cite{PhysRevB.92.041105,dsheng}, here the long-range
dipolar exchanges suppress the $120^\circ$ order to favor a
disordered state. 
Our result agrees with earlier DMRG study of 
dipolar Heisenberg model for $\phi=0$ with truncated interactions \cite{yao2015quantum}. 
Both predict a spin disordered phase for $\theta<\theta_c$, with $\theta_c\sim 19^\circ$
from pf-FRG and $\theta_c\sim 10^\circ$ from DMRG. 
Our new insight is that  the PM phase becomes much broader, $\theta_c\sim 54^\circ$,
if we tune $\phi$ to $30^\circ$ to suppress the stripe phase.

Now we discuss the pf-FRG results for a few representative points on the phase
diagram.
Let us start with the point $\mathcal{P}_1$ in Fig.~\ref{fig:triangular},
$\theta=10^\circ$, $\phi=5^\circ$. The spin susceptibility profile
$\chi(\mathbf{p})$ at $\Lambda\approx 0.2$ is
shown in the middle panel. It peaks at $K$ and $K'$, indicating the
$120^\circ$ correlations. There is however no long-range order. 
We find instead a remarkably smooth flow of $\chi_\mathrm{max}$ down to $\Lambda\rightarrow 0$
without any sign of instability in the bottom panel of
Fig.~\ref{fig:triangular}. Note that small fluctuations at small $\Lambda$
are artifacts due to the frequency discretization and they diminish with finer grid.
Similar PM behaviors are observed for point $\mathcal{P}_2$ and $\mathcal{P}_3$
at larger values of $\phi$, with the $\chi(\mathbf{p})$ profile titled accordingly.
These are our most significant findings.

Moving from point $\mathcal{P}_1$ towards $\mathcal{M}_1$, the 
peaks at $K$ and $K'$ first become flatter and eventually coalesce at 
$\theta=17^\circ$. Here $\chi_\mathrm{max}$ shows a massive degeneracy in $\mathbf{p}$-space:
it peaks along the entire $K$-$K'$ line.  
Beyond this point, the flow of $\chi_\mathrm{max}$ 
shows increasing jumps at small $\Lambda$, and 
a kink (or turning point, indicated by the small arrow) is developed for $\theta> 19^\circ$.
At the point $\mathcal{M}_1$, $\chi(\mathbf{p})$ is sharply peaked at $M$, 
and the flow becomes unstable at small $\Lambda$, clearly indicating the stripe phase. 
Increasing $\theta$ further beyond the point $\mathcal{M}_2$, 
$\chi(\mathbf{p})$ develops a peak at a location between the $M$ and $\Gamma$ point. 
Similar result is obtained for other values of $\phi$, such as the  
 $\mathcal{M}_3$ point  in Fig.~\ref{fig:triangular}. Here, the sharp peak of $\chi(\mathbf{p})$
as well as the unstable flow unambiguously identify the spiral order. 

To locate the phase boundaries in a systematic manner, we introduce an empirical measure
to quantify and detect the breakdown of smooth FRG flow. For a given dipolar tilting, we 
compute $
f(\theta,\phi) = \sum_{\Lambda} (\chi_\mathrm{max}\big|_{\Lambda}
 -\chi_\mathrm{max}\big|_{\Lambda-d\Lambda} )^2 $,
i.e. the ``sum of unphysical jumps'' during the flow. 
The value of $f$ is very small in the paramagnetic phase
because of smooth continuous flow, 
and very large for ordered phases because their unstable flow \cite{supp}. By comparing
to DMRG, we know that at $\theta=0$, independent of $\phi$, the system
is deep inside the PM phase. Thus, it provides a standard measure $f_0=f(\theta=0,\phi=0)$.
If $f(\theta,\phi)\leq f_0$, the low energy flow is equally smooth or even smoother
than that at $\theta=0$, we then conclude the system flows to a disordered, paramagnetic
phase.
The resulting boundary of the PM phase is shown by the solid white line bending to the right in
Fig.~\ref{fig:triangular}. The transition from PM to stripe is marked by a rapid 
increase in $f/f_0$. In contrast, the transition from stripe to spiral is signaled by 
smoothening of the flow and thus suppression of $f$ (see the flow for $\mathcal{M}_2$ in Fig.~\ref{fig:triangular}). 
We identify the stripe-spiral boundary as where $f$ develops
a local minimum along the horizontal cuts on the $(\theta,\phi)$ plane. It is shown by the white dashed line bending to the left
in Fig.~\ref{fig:triangular}. This line is also where the peaks in $\chi(\mathbf{p})$ become smeared and 
the location of $\mathbf{p}_\mathrm{max}$ begins to change character.
In the hatched region around $\mathcal{M}_2$ in Fig.~\ref{fig:triangular}, the flow is much smoother than $\mathcal{M}_1$ and $\mathcal{M}_3$ 
at small $\Lambda$. It is analogous to $\mathcal{P}_1$ but $\chi_\mathrm{max}$ reaches a much
bigger value at $\Lambda=0$. Therefore, this small region is likely a second PM phase, but on the verge
of being ordered. 

To summarize, our numerical FRG calculation reveals a quantum paramagnetic
phase occupying a large portion of the phase diagram of the dipolar
Heisenberg model. FRG enables us to reach large cutoff distances 
for an accurate description of the dipolar exchange and the spiral order. It
describes quantum fluctuations beyond
the spin-wave or Schwinger-boson theory.  The widespread lack of divergence
in $\chi$ is unexpected. In hindsight, three factors conspire to suppress
long-range order. First is the lattice geometric frustration. 
Second, the stripe order is completely suppressed for dipole titling $\phi=30^\circ$ due to symmetry, 
such that the paramagnetic phase extends to as far as $\theta=54^\circ$.
Third is the competition of $J_{ij}$, i.e. exchange frustration, stemming from the 
long-range dipolar exchange (see Fig.~\ref{fig:geometry}). 
Even in the $J_1$-$J_2$ model, finite $J_2$
enhances paramagnetic behavior \cite{dsheng,PhysRevB.92.041105,PhysRevB.96.075117}. Longer range exchanges
lead to very flat
classical energy landscape, with distinct orders close in energy. These weak classical orders are melted by quantum
fluctuations to form a quantum paramagnet.

Our results suggest that experiments on ultracold quantum gases of polar molecules with electric dipole moments 
or atoms with large magnetic dipole moments are promising systems to explore frustrated magnetism and search for spin 
liquids \cite{yan2013observation,PhysRevLett.111.185305}.
There are two limitations to the current pf-FRG method. First, the flow equation is restricted to one-loop diagrams.
An improvement is to include two-loop terms as achieved recently in Ref. \cite{rueck2017effects}.
Second, current pf-FRG implementation cannot directly characterize the spin liquid states.
Future work is needed to elucidate the nature of the 
predicted spin liquid states in various spin $1/2$ models
on the triangular lattice \cite{PhysRevB.87.104406,PhysRevB.93.165113}, which remains an outstanding open problem.

\begin{acknowledgments}
This work is supported by NSF Grant No. PHY-1707484 and
AFOSR Grant No. FA9550-16-1-0006. A.K. is also supported by ARO Grant No. W911NF-11-1-0230.
The GPU used for the calculation is provided in part by the NVIDIA Corporation.
\end{acknowledgments}
\bibliography{refs}

\pagebreak
\widetext
\begin{center}
\textbf{\large Supplementary Materials for ``Absence of long-range order in a triangular spin system with dipolar interactions''}
\end{center}
\begin{center}
  Ahmet Kele\c{s} and Erhai Zhao
\end{center}

\title{}

\setcounter{equation}{0}
\setcounter{figure}{0}
\setcounter{table}{0}
\setcounter{page}{1}
\makeatletter
\renewcommand{\thefigure}{S\arabic{figure}}
\renewcommand{\thetable}{S\arabic{table}}
\renewcommand{\theequation}{S\arabic{equation}}
\makeatother

\section{Classical Phase Diagram}

\begin{figure}[h]
  \centering
  \includegraphics[width=0.5\textwidth]{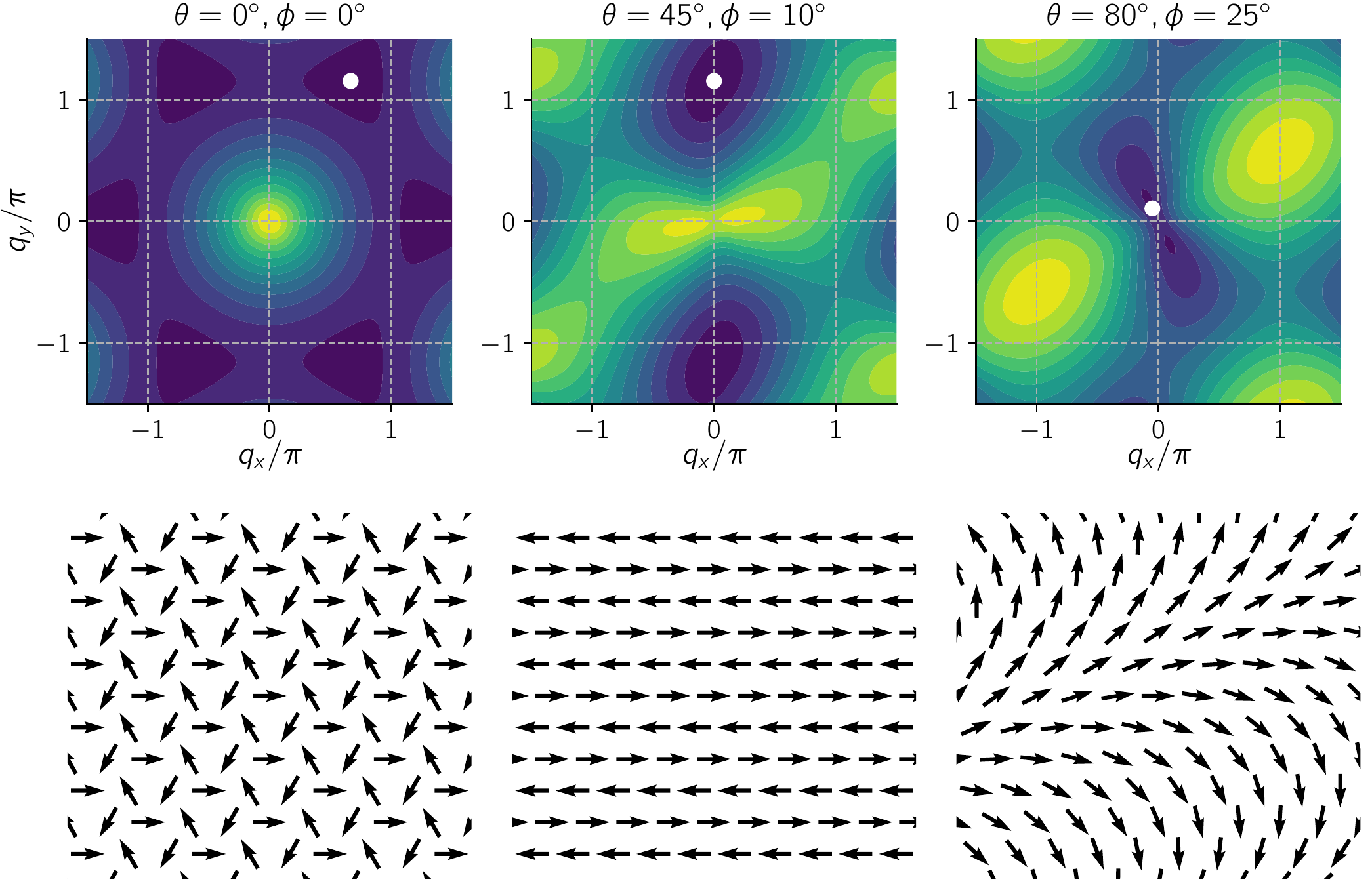}
  \includegraphics[width=0.3\textwidth]{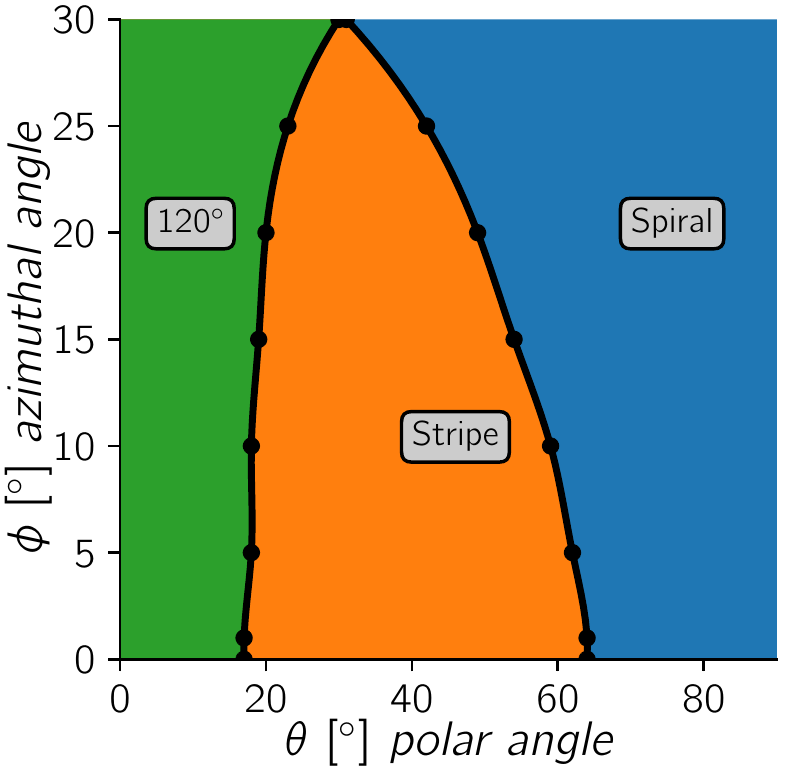}
  \caption{Left: the top row shows classical energy landscapes $E_\mathrm{cl}(\qb)$ as a function of
    $q_x$ and $q_y$ for three different dipolar tilting angles. Dark (blue) regions
    are energy minima whereas bright (yellow) regions are energy maxima.
    The wavevector
    minimizing the energy is indicated with a white dot in each figure. The
    bottom row shows the ordering (spin configuration) in real space for the minimum energy
    wavevector $\qb=\qb_{min}$. Right: the phase diagram obtained from classical analysis
    showing three phases.}
\label{fig:classical_phase_diagram}
\end{figure}

To find the classical ground states of dipolar Heisenberg model, we substitute the operator $\mathbf{S}$ with classical vector
$S_i = S( \cos\qb\cdot\rb_i,\sin\qb\cdot\rb_i, 0 )$ in the Hamiltonian $H=\sum_{ij}J_{ij} \mathbf{S}_i\cdot \mathbf{S}_j$ with dipolar
coupling $J_{ij}$ and obtain the classical energy
\begin{equation}
    E_\mathrm{cl}(\qb)   = S^2\sum_{ij} J_{ij} 
    \cos(\qb\cdot \rb_i-\qb\cdot\rb_j
    ). 
\end{equation}
We minimize $E_\mathrm{cl}$ with respect to $\qb=(q_x,q_y)$ to find the
ordering wave vector $\qb_{min}$. As an
example, we plot the energy landscapes $E_\mathrm{cl}(\qb)$ and the corresponding spin configurations in real
space for three representative angles in Fig.~\ref{fig:classical_phase_diagram}.
Here, we have chosen a 40-by-40 lattice grid and performed
the summation numerically for each value of $\qb$ in the Brillouin zone. By
performing this analysis for each point in the parameter space of dipolar
tilting $(\theta,\phi)$, the phase
diagram of the system can be found as shown in
Fig.~\ref{fig:classical_phase_diagram}. 

From the locations of the minima in the energy landscape, one can see the
$120^\circ$ order for small tilting with ordering vector
$\qb_{min}=2\pi(1/3,1/\sqrt{3})$, the stripe order for intermediate tilting
with $\qb_{min}=2\pi(0,1/\sqrt{3})$, and finally the spiral order with wavevectors
that are generally incommensurate with the lattice spacing and change
continuously within the Brillouin zone with dipolar angles. 
Fig.~\ref{fig:classical_phase_diagram} is obtained by scanning the 
$(\theta,\phi)$ plane along a few horizontal cuts and
finding the phase boundaries along these cuts.
\section{Frequency independent Functional Renormalization} 

When frequency dependence of the vertex is ignored, the self energy term in the
flow equations vanishes and the following 
spin parametrization can used
\begin{equation}
    \Gamma_{1',2';1,2} = \Gamma_s( i_1,i_2 )\delta_{i_1'i_1} \delta_{i_2'i_2}
    \sigma_{\alpha_1'\alpha_1}\sigma_{\alpha_2'\alpha_2} - (1'\leftrightarrow
      2'), 
\end{equation}
which obeys the following simplified flow equation 
\begin{equation}
    \frac{d}{d\Lambda}\Gamma_{s}(i_1,i_2)=
    \frac{2}{\pi}\frac{1}{\Lambda^2} \biggr[
   -2\Gamma_{s}(i_1,i_2) 
     \Gamma_{s}(i_1,i_2) 
   + \Gamma_{s}(i_1,i_2)\Gamma_{s}(i_1,i_1) 
   +\sum_{j} \Gamma_{s}(i_1,j)\Gamma_{s}(j,i_2) \biggr].  \label{s2}
\end{equation}
Note that a parametrization in the density channel is also allowed by symmetry
in the form $\delta_{\alpha_1'\alpha_1}\delta_{\alpha_2'\alpha_2}$, however
such a term vanishes under frequency independent renormalization flow as
demonstrated in Ref.~\cite{Reuther2010} and therefore can be neglected (we will implement parametrization in
the density channel in the next section). The factor in front of the right hand
side of Eq. \eqref{s2} comes from the remaining frequency dependence of Green function bubble
and the integral of this term is calculated as 
\begin{align}
  \frac{1}{2\pi}\int d\w2\Pi(\w,\w) 
     &= \frac{1}{\pi} \int d\w \frac{1}{\w^2 } 
     \delta(|\w-\Lambda|) \Theta(|\w|-\Lambda)   \nonumber\\
     &=\frac{1}{\pi} \left[\frac{1}{2} \frac{1}{\Lambda^2}
       +\frac{1}{2} \frac{1}{(-\Lambda)^2} \right] = \frac{1}{\pi}
     \frac{1}{\Lambda^2 }. 
\end{align}
The first order coupled differential equations are solved
numerically starting from the initial condition at the bare (ultraviolet)  scale $\Lambda_{UV}$
\begin{equation}
    \Gamma_s(i_1,i_2)\biggr|_{\Lambda=\Lambda_{UV} } = \frac{1}{4} J_{ij}  
\end{equation}
down to the infrared scale $\Lambda_0\ll 1$. 
From the vertex, the static spin susceptibility can be calculated as 
\begin{equation}
    \chi( i_1,i_2 ) \equiv \langle S^z_{i_1} S^z_{i_2}\rangle = \frac{1}{2\pi\Lambda} \delta_{i_0i_2} -
    \frac{1}{\pi^2\Lambda^2}\Gamma_s(i_1,i_2).  
\end{equation}

From translational invariance, the vertex $\Gamma_s$ only depends on $i_1-i_2$. To monitor
the flow, we Fourier transform the susceptibility and inspect its momentum
space profile. Since the only energy scale
in the problem is the
exchange energy $J_0$ which is taken to be the unit of energy, we set up
a renormalization grid for $\Lambda$ starting from two orders of magnitude smaller than this
scale $\Lambda_0\propto10^{-2}$ up to two orders of magnitude larger than the
exchange scale $\Lambda_{UV}\propto 10^{2}$ with logarithmically spaced mesh. We have
checked that further increasing this interval (e.g. increasing $\Lambda_{UV}$) does not 
change the FRG flow noticeably.

\begin{figure}[h]
  \centering
  \includegraphics[width=0.5\textwidth]{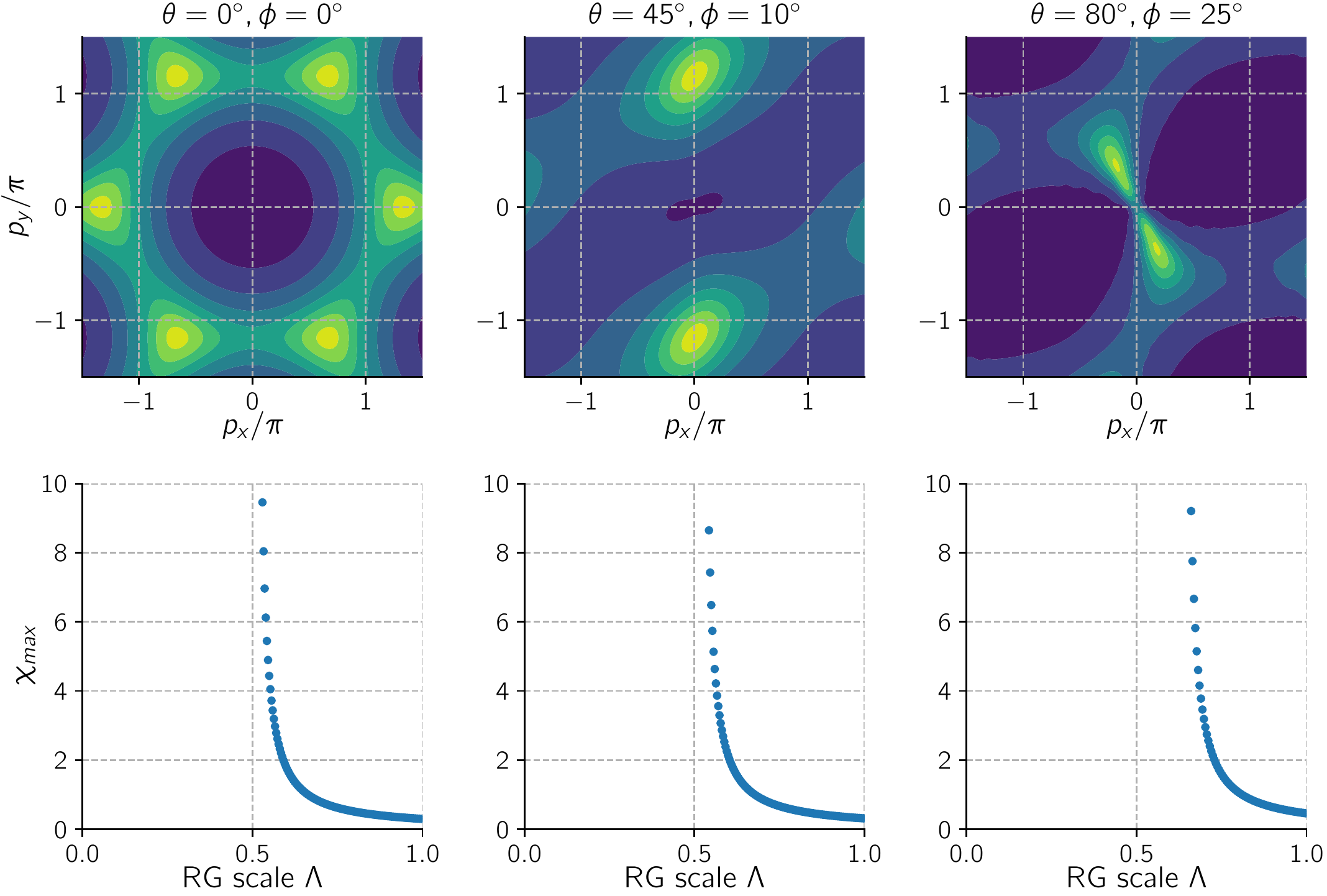}
  \includegraphics[width=0.3\textwidth]{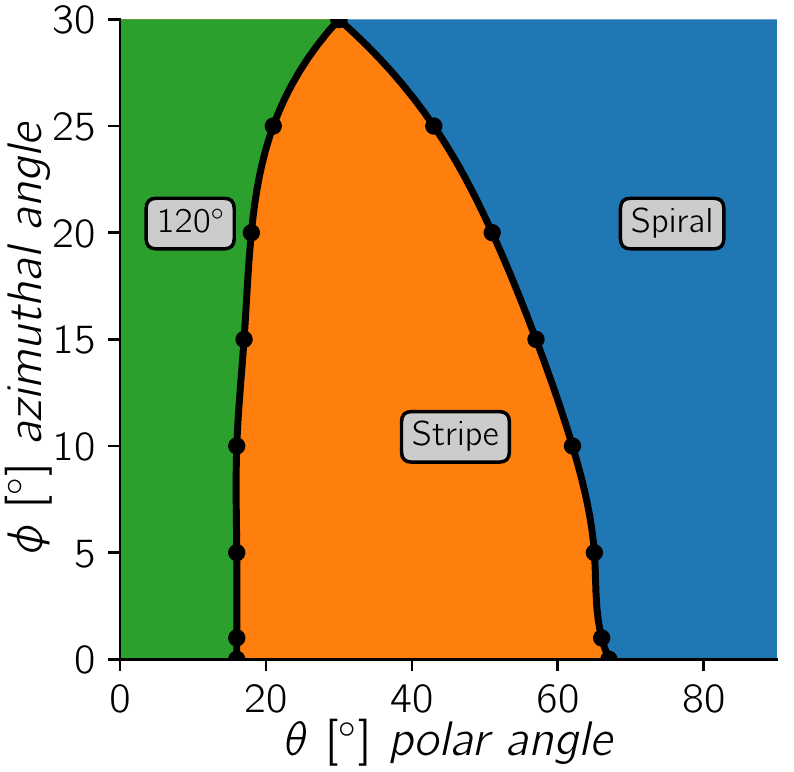}
  \caption{Right: phase diagram of the dipolar Heisenberg model on triangular lattice from static (frequency-independent) FRG
    calculation, in good agreement with classical analysis. Left: the susceptibility profiles for three
    selected points in the phase diagram and the corresponding renormalization flow of
    the susceptibility maximum. The divergence of spin susceptibility maximum indicates long-range order. Dark
    (blue) regions are susceptibility minima whereas bright (yellow) regions
    are susceptibility maxima. }
    \label{fig:static_flows}
\end{figure}

The phase boundary can be located by inspecting the susceptibility profile $\chi(\mathbf{p})$
for each parameter to determine its peak location. Since the sharp peaks turn to flat ridges
near the phase boundaries, it is much more convenient to pinpoint the phase boundary by monitoring the 
degeneracy of the susceptibility profile. For example,
as we move along a horizontal cut going from the $120^\circ$ phase to the stripe phase, the maxima at
$K$ and $K'$ points gets elongated and eventually merge to give the shape of an extended ridge
along the $K-K'$ line at the phase boundary.
Once we enter the stripe phase, the maxima at $M$ point dominates to lift the
degeneracy along the $K-K'$ line. 
To determine the degeneracy, we divide the momentum
space into $N_{bins}$ bins and count $N_{peak}$, the number of bins within which
$\chi(\mathbf{p})$ reaches at least 90 
percent of its maximum value. Then the degeneracy is assigned as $N_{peak}/N_{bins}$.
The phase diagram resulting from this analysis is shown in
Fig.~\ref{fig:static_flows}. It is very similar to the classical diagram Fig. \ref{fig:classical_phase_diagram}.
Note the static FRG takes into quantum fluctuations not considered in the classical analysis.
As a result, the $120^\circ$-stripe boundary is pushed slightly to the left, while the stripe-spiral boundary is shifted
slightly to the right, each by about 2$^\circ$. The agreement with the classical phase diagram serves 
as a sanity check for the back bone of our FRG calculation.

\section{Details of Full Functional renormalization implementation}

For numerical implementation of FRG flow equations with full frequency
dependence, we use the following parametrization
\begin{equation}
    \Gamma_{1',2',1,2} = \Gamma^\mathrm{spin} _{i_1i_2}(s,t,u) \sigma_{\alpha_1'\alpha_1}
    \sigma_{\alpha_2'\alpha_2} \delta_{i_1'i_1} \delta_{i_2'i_2} +
    \Gamma^\mathrm{density} _{i_1i_2}(s,t,u) \delta_{\alpha_1'\alpha_1}
    \delta_{\alpha_2'\alpha_2} \delta_{i_1'i_1} \delta_{i_2'i_2} 
    -(1'\leftrightarrow 2') 
\end{equation}
where $(1'\leftrightarrow 2')$ implements fermionic antisymmetry and frequency
  dependence is expressed in terms of Mandelstam variables
  $s=\w_1+\w_2=\w_1'+\w_2'$, $t=\w_1-\w_1'$ $u=\w_1-\w_2'$. In this
  parametrization, the first term is the spin-spin interaction parametrized by the well known
$\sigma\cdot\sigma$ form, whereas the second term is the density-density
interactions. Site dependence is parametrized by terms like $\delta_{i_1'i_1}$ 
or $\delta_{i_1'i_2}$ so that a $\psi\dgr$ term is matched with $\psi$ term in
the interaction vertex $\psi\dgr\psi\dgr\psi\psi$. This parametrization, along
with site independent self energy and full Green function
\begin{equation}
    G(i_1',\w_1',\alpha_1';i_1,\w_1,\alpha_1;) = \delta_{i_1'i_1} 
    \delta_{\w_1'\w_1}
    \frac{1}{i\w_1+\Sigma(\w_1) } 
\end{equation}
ensures that the fermion hopping is forbidden and number of fermion per site
is fixed at one. This has been discussed in detail in Ref.~\onlinecite{Reuther2010}.
It has been explicitly checked in Ref.~\onlinecite{PhysRevLett.120.057201} that the fermion number
constraint is preserved exactly in pseudofermion FRG at zero temperature as
implemented here.

Once $G$ and $\Gamma$ are obtained from the numerical
renormalization flow, we calculate the static spin susceptibility using the
following diagrammatic expression
\begin{equation}
    \chi_{i_1,i_2} = \int_0^\infty d\tau \langle T S_{i_1}(\tau) S_{i_2}(0)
    \rangle = 
    \begin{tikzpicture}[scale=.5,baseline=(current bounding box.center)]
      
      \draw[densely dashed,very thick] (0,0.7) -- (1,0.7); 
      
      \draw[middlearrow={stealth}] 
      (1,0.7) to [out=75,in=105] (3,.7) ;
      \draw[middlearrow={stealth reversed}] 
      (1,0.7) to [out=-75,in=-105] (3,.7) ;
    
      \draw[densely dashed, very thick] (3,0.7) -- (4,0.7); 
      
      \node[above,black] at (4.5,0.2) {$+$};
      
      \draw[densely dashed, very thick] (5,0.7) -- (6,0.7); 
      
      \draw[middlearrow={stealth}] 
      (6,0.7) to [out=85,in=125] (8,1.0) ;
      \draw[middlearrow={stealth reversed}] 
      (6,0.7) to [out=-85,in=-125] (8,0.4) ;
    
      \draw[fill=gray] (8,0.4) rectangle (8.6,1);
     
      \draw[middlearrow={stealth}] 
      (8.6,1.0) to [out=55,in=95] (10.6,0.7) ;
      \draw[middlearrow={stealth reversed}] 
      (8.6,0.4) to [out=-55,in=-95] (10.6,0.7) ;
     
      \draw[densely dashed, very thick] (10.6,0.7) -- (11.6,0.7); 
    \end{tikzpicture}
\end{equation}
where 
``$\begin{tikzpicture} 
\draw[densely dashed, very thick] (0,0) -- (.5,0); 
\end{tikzpicture} $''
is the representation for the Pauli matrix $\sigma$ for spin at site $i$, 
$S_i=1/2\sigma_{\alpha'\alpha} \psi\dgr_{\alpha'i}\psi_{\alpha i} $.

\subsection{A. Frequency Grid}
We use the zero temperature formalism such that the Matsubara frequencies
become continuous variables $\w\in[-\infty,\infty]$ and frequency summations
become infinite integration $\int_{-\infty}^\infty d\w/2\pi$. 
Since the vertex contains three independent frequency variables (the fourth one
is given by energy conservation), the computational cost increases quickly as the
frequency resolution is increased.  Changing
frequency variables into Mandelstam form significantly reduces the cost
due to symmetry of the vertex $s\rightarrow-s$, $t\rightarrow-t$,
$u\rightarrow-u$ and self energy $\Sigma(-\w)=-\Sigma(\w)$ as shown
in Ref.~\cite{Reuther2010} based on the structure of the flow equations.
As a result, it is sufficient to restrict the frequency
integrations to positive axis $\int_0^\infty d\w$ based on this symmetry. For
numerical implementation, one has to introduce minimum $\w_{min}$ and maximum
$\w_{max}$
frequencies in the positive axis which can be considered as ultraviolet 
and infrared limits of the theory. In choosing these limits, we have to
remember that the only energy scale is the dipolar exchange energy $J_0$
defined
in Eq.~1 of main text which is taken to be the unit of energy in our
theory, i.e. $J_0=1$. Thus, the conditions on the frequency limits are
$w_{max}\gg 1$ and $\w_{min}\ll 1$. In our simulations we take
$\w_{max}=100$ or $400$ and $\w_{min}=0.01$ or $0.005$ and checked that the final results do
not change. 

The most important step in setting up the frequency grid is to decide the number of
points between $\w_{min}$ and $\w_{max}$ and their spacing. 
Recall that the frequency dependence of the running couplings is weak at large frequencies  due
to the Green function bubbles in the flow equation. So we choose a logarithmically
spaced frequency grid, with a spacing increasing with $\omega$, containing $N_\w=48$ or $64$ points 
within the entire frequency range. 
We have check that the final results do not change with further increase in $N_\w$ or
when we switch to a grid with spacing given by a geometric series
$\w_n=\w_{min} (\epsilon^n-1)/(\epsilon-1)$ with grid parameter $\epsilon$
determined by
$\w_{n=N_\w}=\w_{max}$.
Once the frequency grid is set up, the RG grid is automatically determined
which can be symbolically demonstrated as
\begin{equation}
  \includegraphics[width=0.3\textwidth]{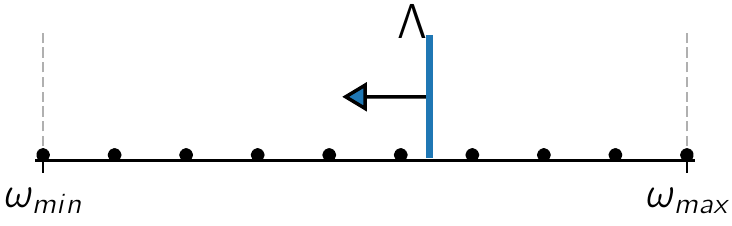} 
  \nonumber
\end{equation}
We take four renormalization steps between each frequency grid and solve the flow
equation using standard Runge-Kutta method. Due to discreteness of
the frequencies, there may be small ``oscillations" in the flow of $\chi$ as $\Lambda$ moves from one
frequency grid point to the next (see for example, point $\mathcal{P}_1$ and $\mathcal{P}_3$
in Fig.  \ref{fig:finite_size}). These small oscillations are natural for any
numerical implementation and well controlled: they can be
reduced by choosing a denser grid.

\subsection{B. Truncation of Interaction Range}

We emphasize that our calculation is {\it not} done for a finite lattice with open boundary conditions.
We assume translational invariance
such that the vertex $\Gamma^\mathrm{spin}_{i_1i_2}$ and
$\Gamma^\mathrm{density}_{i_1i_2}$ only depend on the difference
$\rb_{i_1}-\rb_{i_2}$ which
can be visualized as a bond between site $i_1$ and site $i_2$. 
Note that each bond has three frequency variables. In practice, the computational cost increases
rapidly with the number of bonds and the number of discrete frequencies.
We choose a reference site $i_1$ at the origin, consider
all bonds within the region of a $N_L$-by-$N_L$
parallelogram and discard bonds outside this region. (We have also tried a circular cutoff region with radius $N_L$.
The results are essentially the same for sufficiently large $N_L$. The parallelogram is adopted because it 
is easy to implement on parallel computing platforms.) Formally, this
corresponds to an infinite spin system with truncated effective interaction
range.  (In the
pfFRG literature, this truncation in the interaction range is simply referred to as  ``the cluster size'' for
brevity. It is not to be confused with a finite lattice with open boundary conditions. There is no physical boundary in our implementation. In passing, we mention that periodic
boundary conditions have recently been considered within pfFRG in an
alternative implementation in Ref.~\cite{PhysRevB.85.214406}.) It is critical to choose a sufficiently 
large $N_L$ while keeping the calculation feasible on modern GPU devices.

\begin{figure}[h]
  \centering
  \includegraphics[width=0.8\textwidth]{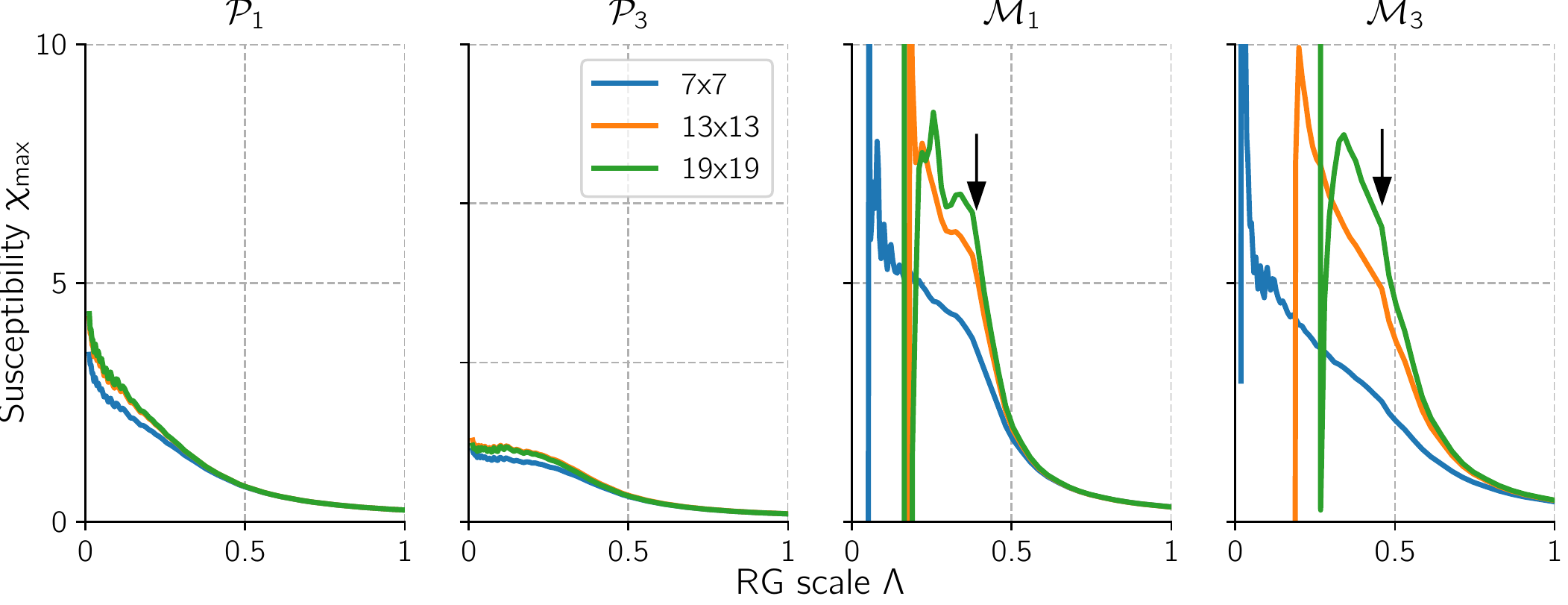}
  \caption{Renormalization flows of the susceptibility maxima $\chi_\mathrm{max}$ toward paramagnetic
    phases ($\mathcal{P}_1$, $\mathcal{P}_3$) and magnetic
    phases ($\mathcal{M}_1$, $\mathcal{M}_3$) for different truncations of
    interaction range $N_L=7, 13$ and $19$. 
The black arrow indicates the critical scale $\Lambda_c$.
}
  \label{fig:finite_size}
\end{figure}

From the real space susceptibility data 
$\chi_{i_1,i_2}\equiv\chi( \rb_{i_1}-\rb_{i_2} )$,
we take the Fourier transform  as
\begin{equation}
    \chi(\pb) = 
    \sum_{i_1-i_2} \chi_{i_1,i_2} 
    e^{i\pb\cdot(\rb_{i_1} -\rb_{i_2}) }.
\end{equation}
Here, the number of points $\pb$ in the Brillouin zone can be taken as arbitrarily
large, since an infinite lattice is assumed. 
By inspecting $\chi(\pb)$, we determine the leading ordering
tendency from the maximum susceptibility $\chi_\mathrm{max}$ and monitor its RG flow.
To understand the susceptibility flows, one can view heuristically 
the renormalization scale $\Lambda$ as the temperature of
the system. Probing collective phenomena at lower energies by reducing the sliding RG scale
corresponds to reducing the temperature of the system. At
low temperatures, the system either enters into a magnetically ordered
phase or stays paramagnetic down to the lowest numerical scales.
If a magnetic order develops, the correlations functions such as the spin
susceptibility would show
divergence at some critical temperature. At this point, the correlation length becomes infinite 
indicating long range order in the thermodynamic limit. However, since we have to truncate the correlations
due to finite computational resources, such divergences will be regulated. Therefore, instead
of a simple divergence commonly encountered in single channel RG, the growth of  $\chi_\mathrm{max}$
beyond the critical scale $\Lambda_c$ will eventually be replaced by unstable flows as shown in Fig.~\ref{fig:finite_size} for point $M_1$ and $M_3$.
This is the price we have to pay for retaining all channels on equal footing and flowing the entire $\Gamma$ instead
of just a few running couplings. In practice, the breakdown of smooth flow beyond the critical scale is actually a blessing. 
It serves as a telling sign of the many-body instability toward the development of long range order.
In contrast, if the FRG flow smoothly reaches the lowest infrared scale without any disturbance (see point $\mathcal{P}_1$ and $\mathcal{P}_3$
in Fig.~\ref{fig:finite_size}), there is no indication
for long range order and the ground state is a quantum paramagnet.

We have systematically investigated the effect of varying $N_L$ on the flow and phase boundary.
Some examples are shown in Fig. \ref{fig:finite_size}  for a few different 
truncations ($N_L=7$, $13$ and $19$).

\textit{Paramagnetic phase --} We observe that the
RG flows for parameters within the quantum paramagnetic phase, such as point $\mathcal{P}_1$ and $\mathcal{P}_3$, 
remain smooth and finite. Susceptibilities stay almost unchanged when we double or triple $N_L$. Such $N_L$-independence
points to a quantum paramagnetic phase in the thermodynamic limit, and is consistent with
the conjecture that it is a spin liquid with short range correlations. 

\textit{Magnetic phases --} 
On the other hand, for 
the stripe ($\mathcal{M}_1$ point) and spiral phase
($\mathcal{M}_3$ point), the onset of the long range order 
can be easily identified from the RG flows in
Fig.~\ref{fig:finite_size}. Here the susceptibility shows a divergence tendency
until a critical scale $\Lambda_c$ which is indicated roughly by the black arrows.  For these two points, the divergence becomes stronger and the value of 
 $\Lambda_c$ converges as $N_L$ is increased. Thus, the instability toward long range
order is unequivocal. Below the critical scale
$\Lambda<\Lambda_c$, the
RG flows break down and show unphysical, discontinuous jumps depending on $N_L$, in sharp contrast to 
the continuous flows for point $\mathcal{P}_1$ and $\mathcal{P}_3$.

The spiral phase presents a challenge for FRG calculations. Ideally, a very large $N_L$ is preferred because
the ordering wave vector $\mathbf{p}_\mathrm{max}$ may, in principle, depend on the details of the bare and
effective interactions at longer distances beyond the cutoff $N_L$. This turns out not to be the case. We have checked that 
$\mathbf{p}_\mathrm{max}$ is not very sensitive to $N_L$ provided it is large enough (e.g. $N_L=13$). 
To get an accurate estimation of $\Lambda_c$ and the critical temperature for the spiral phase, which is not our main goal here, 
one has to carefully track the dependence of the flows as $N_L$ is increased, preferably to very large $N_L$,
because the convergence of the flow is slower compared to the stripe phase.
In summary, we emphasize that whether or not the systems flows into a long range ordered phase, as
reflected by the breakdown of the smooth flow, does 
not depend in the choice of truncation $N_L$ (see Fig. \ref{fig:finite_size}). The identification of 
the spiral and stripe phase is thus unambiguous. 

\subsection{C. Estimation of the Phase Boundaries} 

Here we present additional details about the $f$ function introduced to estimate 
the phase boundary between the paramagnetic and
magnetic phases,
\begin{equation}
    f(\theta,\phi) = \sum_{\Lambda} (\chi_\mathrm{max}\big|_{\Lambda}
    -\chi_\mathrm{max}\big|_{\Lambda-d\Lambda} )^2 .
\end{equation}
By comparing the flows in Fig.~\ref{fig:finite_size}, one can see that 
the value of $f$ is very small in the paramagnetic phase because of the smooth
continuous flow, and very large for ordered phases because the unphysical discontinuous jumps below $\Lambda_c$.
The value of $f$ changes dramatically as the phase
boundary is crossed. 
%

\begin{figure}[h]
  \centering
  \includegraphics[width=0.3\textwidth]{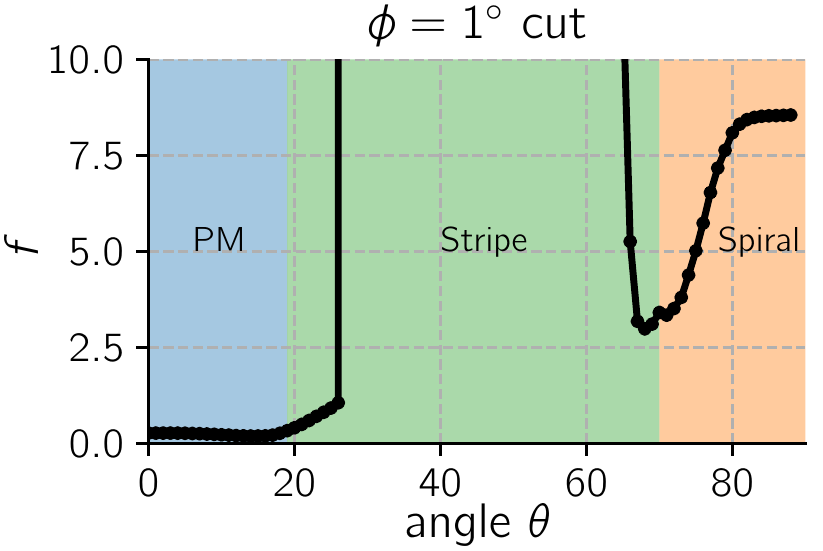}
  \includegraphics[width=0.3\textwidth]{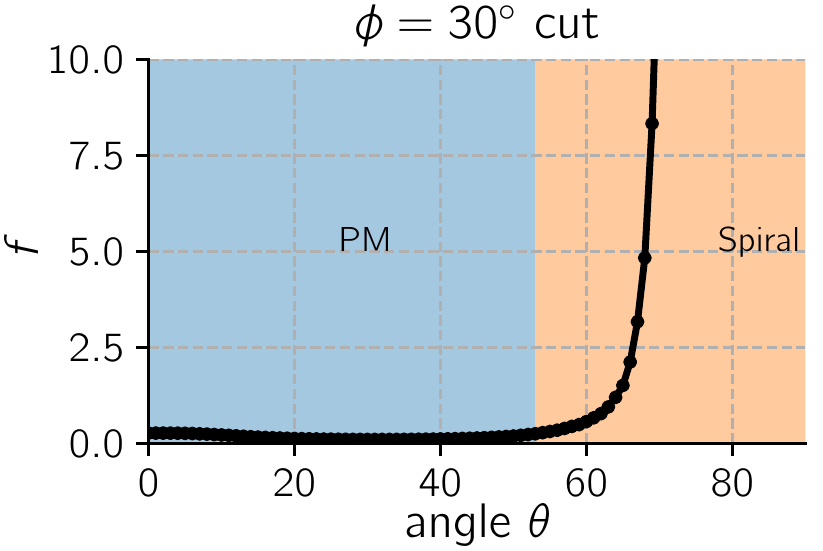}
  \caption{The value of
    $f$ along two distinct horizontal cuts of the phase diagram, 
    $\phi=1^\circ$ (left) and $\phi=30$ (right). }
  \label{fig:phase_boundary}
\end{figure}

\begin{figure}[h]
  \centering
  \includegraphics[width=\textwidth]{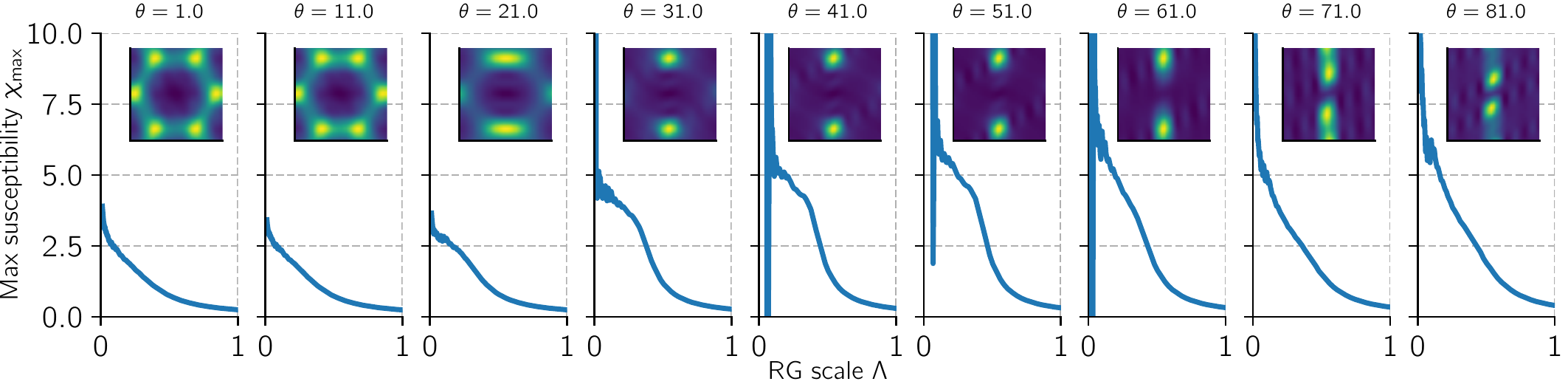}
  \includegraphics[width=\textwidth]{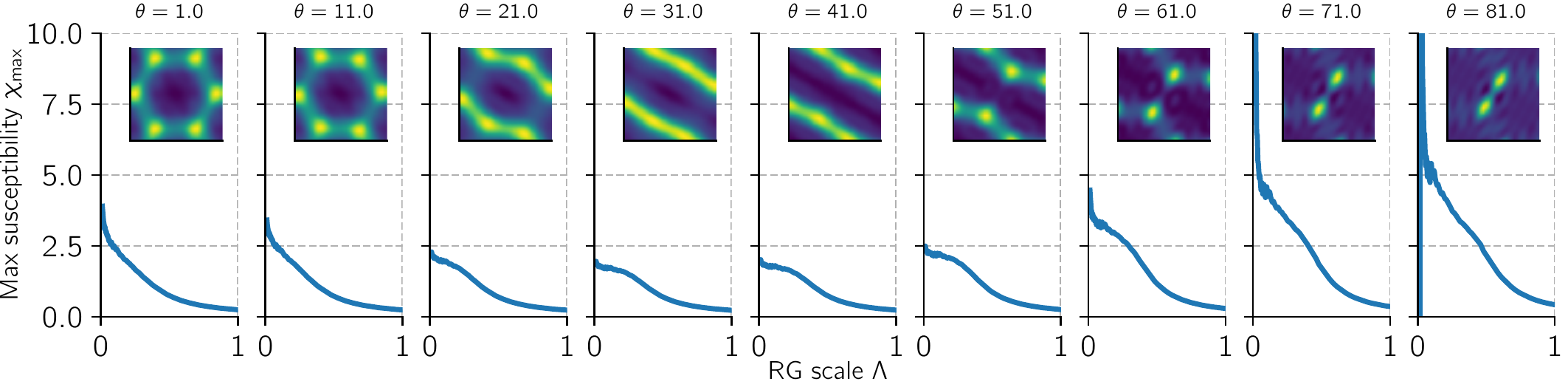}
  \caption{Details of the susceptibility flow for 
    $\phi=1^\circ$ (top row) and $\phi=30^\circ$ (bottom row). The values of $\theta$ are 
    shown in the title of each subplot. Insets
show the susceptibility profiles in the Brillouin zone.}
  \label{fig:phase_boundary_flow}
\end{figure}

As an example, the left panel of Fig.~\ref{fig:phase_boundary} shows the values of $f$
along a cut
for small azimuthal angle $\phi=1^\circ$  and $\theta$ from
$0^\circ$ to $90^\circ$. Details of the susceptibility flow for
selected points along this cut is shown in the top row of
Fig.~\ref{fig:phase_boundary_flow}.  The smooth,
continuous flow at $\theta=0$ indicates a paramagnetic phase. This is consistent with the recent DMRG
study that claims the ground state at this point is a spin liquid \cite{yao2015quantum}.  As $\theta$ is increased, 
$f$ stays flat before dipping slightly around $\theta=16^\circ$ and then starts
increasing slowly afterwards. At $\theta=19^\circ$, $f$ reaches the same value at
$\theta=0^\circ$ and we mark this point as the phase boundary
(the phases are indicated by different color fillings in Fig.~\ref{fig:phase_boundary}). Note that
this is a conservative estimation, because a sharp increase of $f$ occurs later. So the paramagnetic phase may actually persist to larger
values of $\theta$.
In the stripe phase, $f$ is very large. Around $\theta=68^\circ$, $f$ shows
a local minimum. This point is identified as the boundary between the stripe and spiral phase,
and it is consistent with the change in the profile of $\chi_\mathrm{max}(\mathbf{p})$. 


The right panel of Fig.~\ref{fig:phase_boundary} shows the value of $f$
for  azimuthal angle $\phi=30^\circ$. The detailed flows of selected
points are shown at the bottom row of Fig.~\ref{fig:phase_boundary_flow}.
Here the paramagnetic phase persists to much larger $\theta$
angles. The same criterion for $f$ gives the phase boundary at $\theta_c=54^\circ$. The situation here differs from the small
$\phi$ case. The stripe phase is completely suppressed and the paramagnetic
phase directly transitions to the spiral phase.

\end{document}